# Letter/Comment on: Improved predictions for superconductors

Dale R. Harshman [1] and Anthony T. Fiory [2]

[1] Department of Physics, The College of William and Mary, Williamsburg, VA 23187, USA
[2] Bell Labs Retired, Summit, NJ 07901, USA

**Abstract**

Letter/comment on M. Rini, Physics **13**, s94 (July 27, 2020) and A. Sanna *et al*., Phys. Rev. Lett. **125**, 057001 (2020).

With novel materials closing in on superconductivity at room temperature, our interest is, indeed, sparked by the Physics Synopsis by Matteo Rini [1], recounting improved superconducting density-functional theory (SCDFT) reported in [2] and noting the <20% accuracy. Predictions of the superconducting transition temperature $T_C$ for $H_3S$ (at 165 K for denoted "$SH_3$") in [2] and for $LaH_{10}$ included in [2] via citation [3] become especially prescient, given the record high $T_C$ values measured for compressed hydrides. While somewhat improved, predictions based on SCDFT persistently underestimate the actual $T_C$ of $LaH_{10}$, as shown in Fig. 1, which compares pressure dependences of experimental data and theoretical predictions. The black symbols are experimental data (circles from [4], triangles from [5]). As can be seen, predictions from improved SCDFT [3], shown by green square symbols, systematically underestimate experiment, falling 6–14% below data in [4] and 12–17% below data in [5]. Filled red squares ($T_{C0}$) show predictions from electronic mediation at pressures of experimentally maximal $T_C$ in $LaH_{10}$ [6], falling within 1% of experiment, similar to the accuracy for 200-K phase $H_3S$ [6, 7].

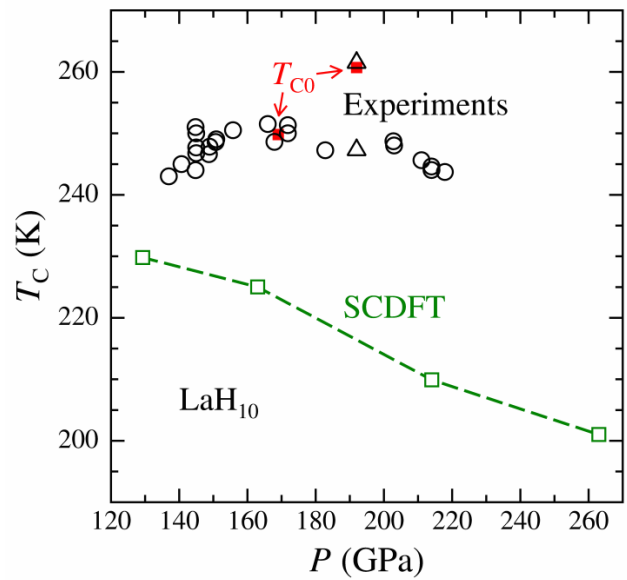

Fig. 1. Experimental and theoretical superconducting critical temperature $T_C$ *vs*. pressure $P$ for $LaH_{10}$.

Clearly, improvement in current phonon theory [3] for $LaH_{10}$, ascribed as preliminary in [2], takes on greater urgency. Once accomplished, it will likely prompt a reevaluation of the theoretical and experimental landscape for superconductivity near room temperature.